# Prototype of an affordable pressure-controlled emergency mechanical ventilator for COVID-19

Américo Pereira[1], Luís Lopes[1], Paulo Fonte[1,2,3,*], Pedro Póvoa[4,5,6], Telmo G. Santos[7], Alberto Martinho[7], Ângela Neves[2], António Bugalho[5,8], António Gabriel-Santos[7], Gonçalo Gaspar Bentes Pimenta[9], João Goês[10], João Martins[10], João Pedro Oliveira[10], José Paulo Santos[11], Luís C. Gil[7], Miguel Onofre Domingues[12], Orlando Cunha[1], Pedro Pinheiro de Sousa[13], Tiago A. Rodrigues[7], Valdemar R. Duarte[7], Antero Abrunhosa[2], António Grilo[7], João Agostinho do Nascimento[14,15], Mário Pimenta[1], for the Project Open Air[15]

[1] LIP – Laboratory of Instrumentation and Experimental Particle Physics, Coimbra, Portugal
[2] ICNAS Institute of Nuclear Science Applied to Health, University of Coimbra, Coimbra, Portugal
[3] Coimbra Polytechnic – ISEC, Coimbra, Portugal
[4] Polyvalent Intensive Care Unit, Hospital de São Francisco Xavier, CHLO, Lisbon, Portugal
[5] NOVA Medical School, New University of Lisbon, Portugal
[6] Center for Clinical Epidemiology and Research Unit of Clinical Epidemiology, OUH Odense University Hospital, Denmark.
[7] UNIDEMI, Department of Mechanical and Industrial Engineering, NOVA School of Science and Technology, Universidade NOVA de Lisboa, 2829-516 Caparica, Portugal
[8] Pulmonologist, CUF Hospitals Lisbon;
[9] Mechanical Design Engineer, Milton Keynes, United Kingdom
[10] DEEC/FCT/UNL, CTS/UNINOVA, 2829-516 Caparica, Portugal
[11] Laboratory of Instrumentation, Biomedical Engineering and Radiation Physics (LIBPhys-UNL), Department of Physics, NOVA School of Science and Technology, NOVA University Lisbon, 2829-516 Caparica, Portugal
[12] General Surgeon, Military Doctor, Portuguese Army
[13] Head of Structural Design, Haas F1 Team, Maranello, Itália
[14] Harvard University, Cambridge, Massachusetts - USA
[15] https://www.projectopenair.org/

## Abstract

We present a viable prototype of a simple mechanical ventilator intended as a last resort to ventilate COVID-19 patients. The prototype implements the pressure-controlled continuous mandatory ventilation mode (PC-CMV) with settable breathing rates, inspiration/expiration time ratios and $FiO_2$ modulation.

Although safe, the design aims to minimize the use of technical components and those used are common in industry, so its construction may be possible in times of logistical shortage or disruption or in areas with reduced access to technical materials and at a moderate cost, affordable to lower income countries. Most of the device can be manufactured by modest technical means and construction plans are provided.

---

[*] Corresponding author: fonte@coimbra.lip.pt.

## Introduction

Coronavirus disease 2019 (COVID-19) is the result of an infection caused by severe acute respiratory syndrome coronavirus-2 (SARS-CoV-2). Clinical presentation ranges from mild respiratory tract symptoms to acute respiratory distress syndrome and sepsis, which can be lethal. Of those infected, there is an important percentage of patients that develops severe SARS-CoV-2 pneumonia and require early invasive mechanical ventilation.

Further than the individual impact, COVID-19 has major consequences for national healthcare systems. The evolution and extent of the outbreak causes enormous pressure and increasing demand for ventilation equipment that may prevent high mortality rates. The number of patients needing invasive mechanical ventilation has surpassed in several high-income countries the installed capacity and the existing manufacturers are having difficulties to address this increased demand, putting at risk patient's survival.

Additionally, the international procurement of the technical materials and components necessary to produce mechanical ventilators by other entities within the relevant timeframe and in the quantities required is likely to be difficult or impossible. It is then important that an emergency invasive mechanical ventilator design emphasizes simplicity and local availability of components and manufacture, while simultaneously providing an effective and safe ventilatory support.

On the other hand, medical mechanical ventilators are normally sophisticated machines for general use, although this level of sophistication is not needed to save lives. Indeed, several governments have issued guidelines for simplified mechanical ventilators for COVID-19.

Our team of medical doctors, engineers and researchers identified [1] a ventilation mode that is at the simultaneous effective and safe for COVID-19 emergency clinical intensive care as well as technically easy to implement, even considered the practical constraints mentioned above. Such mode is the pressure-controlled continuous mandatory ventilation mode (PC-CMV).

Following our proof-of-concept [1], this paper describes a viable prototype of a PC-CMV mode emergency ventilator that implements the following main characteristics, considered by us as minimal but sufficient for the critical care of COVID-19 patients:

– Positive Inspiratory Pressure (PIP) adjustable in the range 5 to 40 $cmH_2O$

– Positive End Expiratory Pressure (PEEP) adjustable in the range 0 to 20 $cmH_2O$

– Fraction of inspired oxygen ($FiO_2$) modulable between approx. 50% and 100%

– Safety pressure relief valve in the inspiration tube adjustable in the range 0 to 45 $cmH_2O$

– Breathing rate adjustable in the range 12 to 25 breaths per minute (bpm)

– Inspiration/expiration time ratio (I/E) adjustable in the range 1:2 to 1:3

– Optionally, low and high PIP and PEEP alarms

Only locally-sourced industrial components were used, likely available in large quantities worldwide. To reduce dependencies, no consumables are needed except for the respiratory circuits, HME filters and high efficiency filters.

Although all components proposed are robust industrial devices and the design is safe in the sense that it is of extreme technical simplicity and very unlikely to apply a dangerous overpressure to the patient, the device is not a certified medical device and should be considered only as a last resort solution, to be evaluated in light of the local regulations.

## Principle of operation

The schematic representation of the proposed emergency ventilator is shown in Figure 1.

Pure oxygen at the standard pressure of 4 bar (400 kPa) is fed from the hospital supply to an adjustable pressure regulator with output range of 5 to 40 mbar, allowing the PIP pressure to be set by just turning a knob.

The regulator output is fed to the inspiration electrovalve V1. This valve should have enough aperture for the air to pass through easily at normal breathing flows.

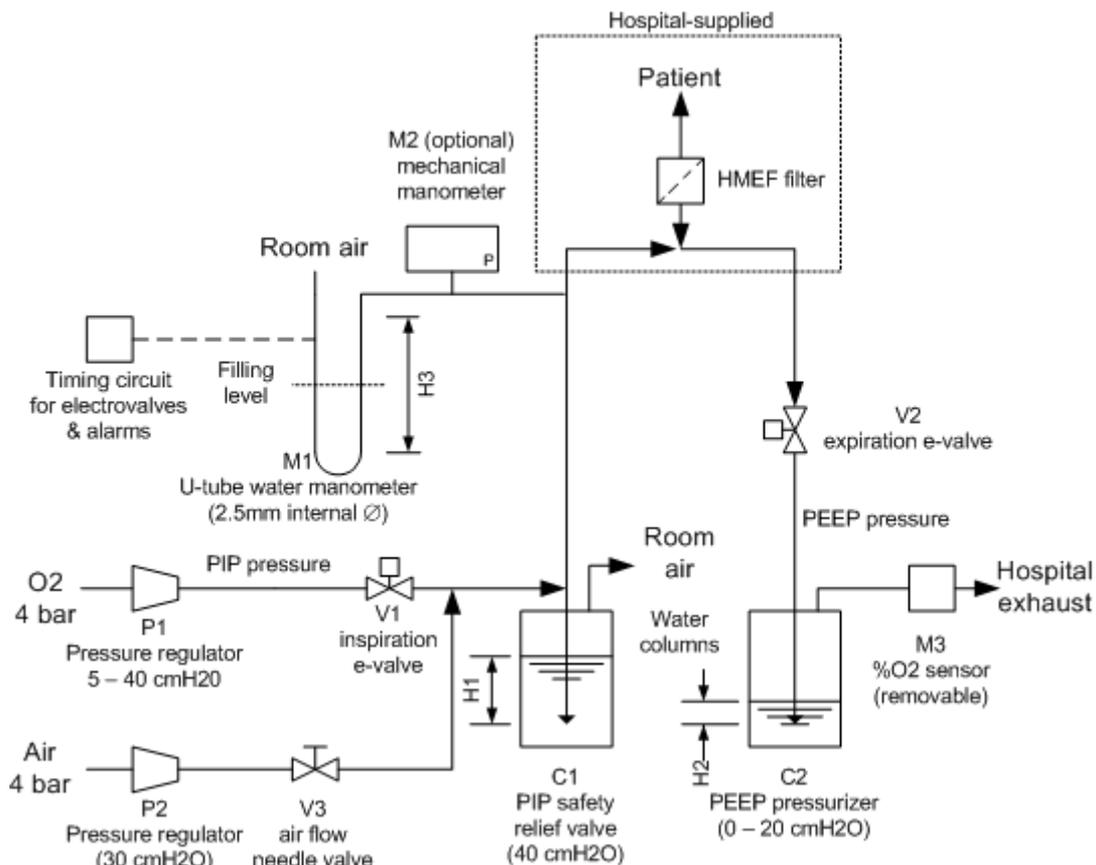

Figure 1 – Schematic representation of the proposed emergency ventilator.

The output of V1 connects to the C1 safety water column, to the M1 water manometer and to the patient inspiration tube, which is a consumable item normally provided by the hospital. Optionally a mechanical manometer, M2, may be applied as well for convenience of measurement.

The C1 water column provides a safety purge to the atmosphere in case of any malfunction that may cause a dangerous overpressure to the patient. It can be set by adjusting the water level H1 to be slightly above the maximum intended PIP. In normal operation there should be no gas flowing through this column. To fulfill its safety purpose, the connecting and inner tubes should have sufficient diameter for a considerable gas flow to pass through without overpressure.

The M1 U-tube water manometer measures the PIP and PEEP pressures directly in $cmH_2O$ by mere observation of the water height difference H3. The tube length should comfortably exceed the intended maximum PIP. A tube of 2.5 mm internal diameter filled with plain water was found to provide a dynamic behavior of the water column close to critical damping, optimizing the speed of response to pressure changes while minimizing the oscillations, improving the overall accuracy of the pressure measurement.

The sustained presence or absence of the water at certain levels can be sensed capacitively by electrodes placed close to the tube, enabling the implementation of high and low PIP and PEEP alarms. In the technical annex it is described an alarm of missing oscillation in the water manometer, indicative of an apnea situation.

From the patient "Y-piece" the expiration tube connects to the expiration electrovalve V2, similar to V1.

The expiration air is vented to the atmosphere through the water column C2. The water level H2 directly determines the PEEP pressure. To avoid aerosolizing contaminated water into the room, the exhaust from this column should be directed to the hospital's general exhaust, but with an input from the room air to avoid strong pressure variations in the column, which may affect PEEP.

The exhaust air can be sensed by a permanent or removable $O_2$ fraction sensor, as a proxy to $FiO_2$. We performed measurements both at the exhaust and at the bottom of the test "lung" and found that they match within ±5%.

Medical air at 4 bar can be taken from the hospital's supply and fed to the pressure reducer P2, adjusted either to a safe fixed output pressure of 30 $cmH_2O$ or to the PIP pressure, whichever is higher. The flow of air is controlled by the needle valve V3, for which there is no stringent specification except that it should allow for sufficient flow (in the tens of L/min) to pass through. The maximum admixture of air is limited by the increase it causes on the PIP pressure, which is limited by P2 to be within the safe pressure range. In these conditions we could modulate the $O_2$ fraction generically between less than 60% and 100% (see the Results section) without any significant increase in PIP.

The valves are electrically commanded with adjustable rate and I/E ratio (duty-cycle) in the ranges stated in the introduction. There are several electronic solutions for this functionality, adjustable to the availability of components.

To improve patient safety, apnea and power-loss alarms were implemented.

However, all the electronics involved is simple, analog, and can be made out of very common components, as exemplified in the technical annex.

A note is in order here to justify the extensive use of water columns in this design.

The water column is a component that has many interesting flow-control characteristics: (a) it has no moving parts except the water itself, so it is very reliable and easy to build; (b) it regulates air pressure in the tens of $cmH_2O$ range accurately, adjustably, and quite independently of the flow; (c) if made of transparent materials the observation of the difference in water levels between the two vessels provides direct information of the differential pressure; and (d) it provides check-valve functionality up to a certain reverse pressure.

Besides being bulky, the main drawback of the water column is the water itself, as its level must be adjusted/monitored and may become bacterially contaminated, requiring anti-bacterial treatment. The water may be replaced by other suitable substances such as low viscosity oils.

It is clear that the functionality provided by the water columns can be performed more conveniently by traditional pneumatic components, such as disposable PEEP valves, if available.

## Experimental setup

A general view of the prototype is shown in Figure 2, while details can be seen in Figure 3.

For the regulators we used an adjustable type normally used as the final regulator in gas distribution installations in buildings, with nominal output pressure of 37 mbar [2]. It was found to be adjustable quite precisely from 20 to 40 mbar and that it could accept input pressures as low as 1 bar without perceivable loss of flow. These are by no means specific for this task and many similar devices are likely to be suitable. In the technical annex are presented comparative tests of 5 different models, all yielding similar performances.

For both the V1 and V2 electrovalves we used components normally used for safety reasons in gas distribution installations for buildings and industry [3]. Although we didn't do it that way, it would be advisable that the expiration valve would be normally open and the inspiration valve normally closed.

Other low-pressure valves (e.g. water valves) may be suitable for the purpose, provided they have enough aperture for the air to pass through easily at normal breathing flows. However, their use in pure $O_2$ may raise concerns about the eventual shedding of sparks by the internal mechanism. In this regard the use of components made for flammable gases is more adequate.

High-pressure valves, such as those used for water intake in domestic appliances, may be unsuited as they are often commanded via an internal servomechanism that requires a few bars of pressure to operate. Eventually, such high-pressure valves could be used on the oxygen and air lines before the regulators, but this was not explored. Electrovalves for pneumatics would also certainly work, if ones with sufficient aperture can be found.

The M1 U-tube water manometer was made with a flexible 4 mm external diameter tube (2.5 mm internal) filled with ordinary tap water.

The C1 and C2 water columns were manufactured from a standard acrylic pipe closed by POM machined plugs, so that they can be disassembled for cleaning and sterilization. Drawings can be found in the technical annex. The inner pipe is a standard PVC pipe with 20mm external diameter. A tight (commercial) plastic gasket allows these pipes to slide up and down with some friction, allowing the easy adjustment of the safety PIP and PEEP.

The flexible tubes connecting to the water columns are standard intubation pipes available from medical suppliers. However, as the inhaled air doesn't flow through these tubes/columns the material safety requirements are lower here.

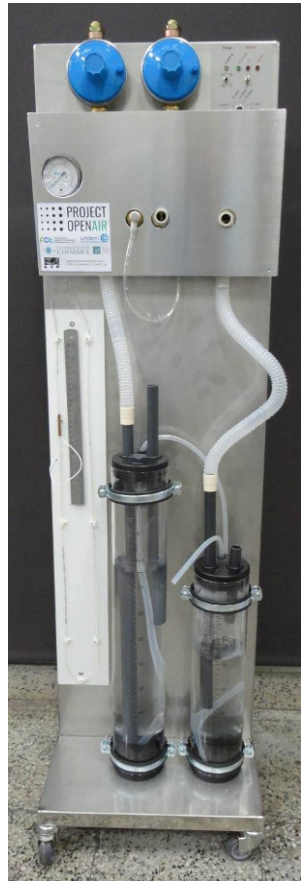

Figure 2 - Global view of the prototype

The flow of air through C2 is somewhat turbulent and reduces the PEEP by about 2 cmH$_2$O with respect to the water height. To alleviate the problem we used a diffuser made of a slightly wider pipe inserted into the PVC pipe, tamponed at the end and with many lateral orifices.

The connecting accessories were standard gas accessories made of brass (generally regarded as a health-safe material) and the tubes connecting to the test "lung" were the actual tubes used in hospitals for this purpose.

The V3 needle valve was a model from the manufacturer SMC, but many other types can certainly be used, as this is a quite undemanding component.

The device was conceived to be mechanically simple and fast to manufacture, with easy access to the water columns for visual verifications and adjustment of the tube and water heights.

The expiration branch needs to be disassembled (by unscrewing four screws) and sterilized between patients. For this a bath of isopropyl alcool (IPA) or ethyl alcool at 70% v/v is recommended.

A test "lung" was made from two sturdy plastic bags meant for urine collection in bedridden patients, each with 1.5 L capacity (Figure 4 – right hand side). The bags were squeezed between two plates forced together by long elastic ropes. The compliance of the "lung" could be adjusted by varying the strength of the ropes.

All components were sourced from local (Coimbra, Portugal) retailers.

The command of the (220 VAC) electrovalves is installed on a standard electrical junction box in the back of the support plate and it has only two commands: a rotary switch with the seven positions "expiration", "inspiration", "12 bpm", "15 bpm", "18 bpm", "21 bpm" and "25 bpm". A switch selects I/E equal to 1/2 or 1/3. The schematic is given in the Technical Annex. The electrovalves themselves are water-tight (IP65) and the metallic frame was properly grounded.

An apnea alarm is implemented by detecting capacitively the movement of the water in the water manometer. If the movement ceases, indicating the absence of a pressure wave, that is, apnea, an audible and visual alarm will be activated after 15s. The audible part of the alarm can be turned off, with indicator.

A video of the device in action can be seen in [4].

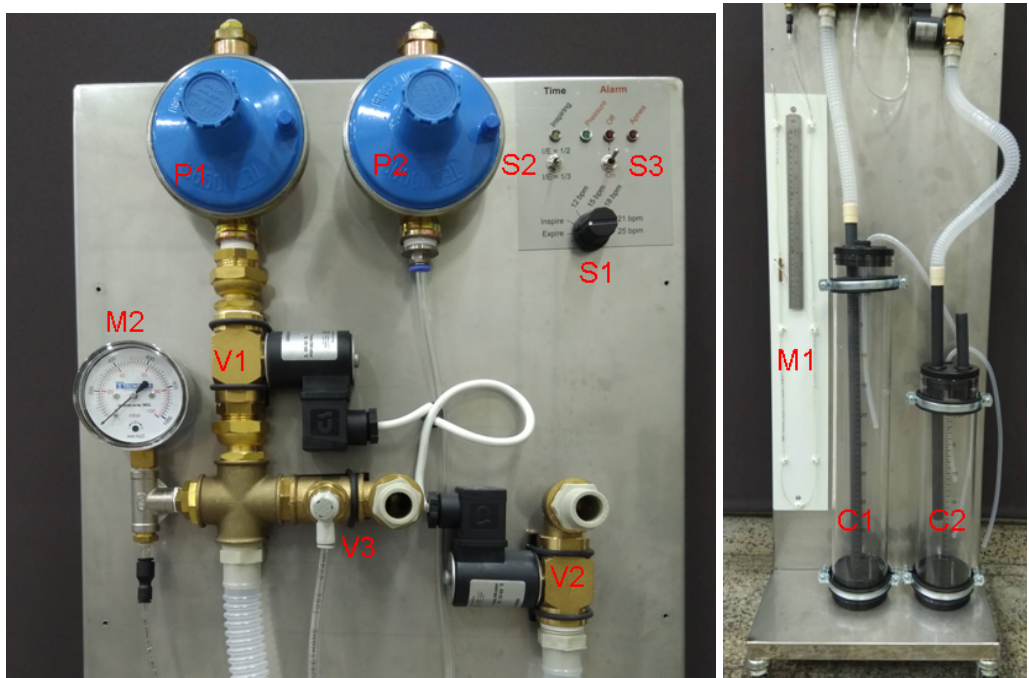

Figure 3 – a) Upper part of the device with the cover removed, indicating the correspondence between the physical objects and the schematic drawing (Figure 1). Refer to the Technical Annex for the electrical switches and LEDs. b) Idem, for the lower part of the device. The capacitive sensor for the apnoea alarm is visible on left branch of M1.

## Results

The prototype was connected to a test "lung" via the tubes used normally for patient intubation, including the relevant filters, as shown in Figure 4. These tubes were kept fully extended.

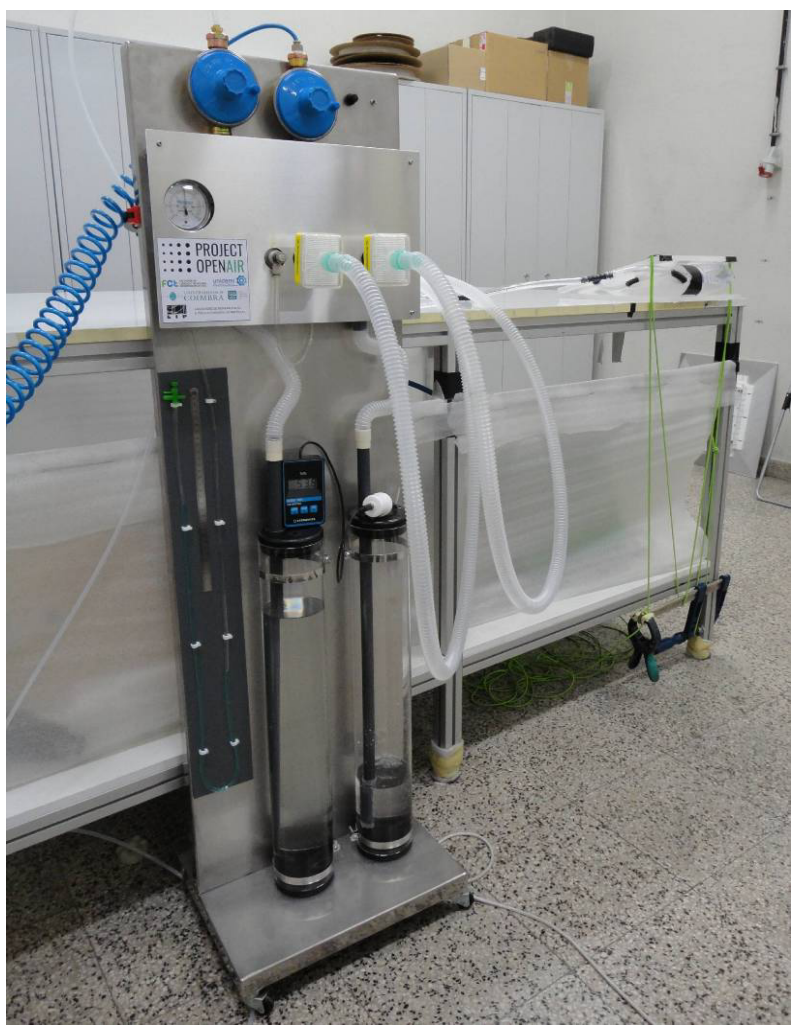

Figure 4 – A view of the test setup. The O2 sensor is visible on top of the C1 safety water column, sensing the gas exhaust from C2.

The test "lung" was calibrated by filling it with a volumetric pump with a displacement of 150 mL. The resulting curves can be seen in Figure 5. By counting the number of steps in the interesting pressure range and the corresponding pressure variation two compliance values were defined, 14 and 32 mL/cmH$_2$O, close to the observed limits for COVID-19 patients.

Figure 6 shows the pressure profiles at the entrance of the "Y piece", measured with electronic manometers as a function of time for breathing rates of approx. 12, 18 and 25 bpm, I/E ratios of 1/2 or 1/3, into a "lung" of compliance 32 mL/cmH$_2$O. The limit PIP (achievable in long inspirations) was set to 20 or 30 cmH$_2$O, compatible with the maximum safe PIP to be considered for the majority of the patients and to the 6 mL/kg maximum expectable volume intake, which was estimated from the calibration and can be read on the right hand side scales.

For all measurements PEEP was set to 12 cmH$_2$O, within the range of advisable values for COVID-19. The regulator P2 was set to 30 cmH$_2$O, except when PIP was set to 40 cmH$_2$O in which case it was set to the PIP pressure.

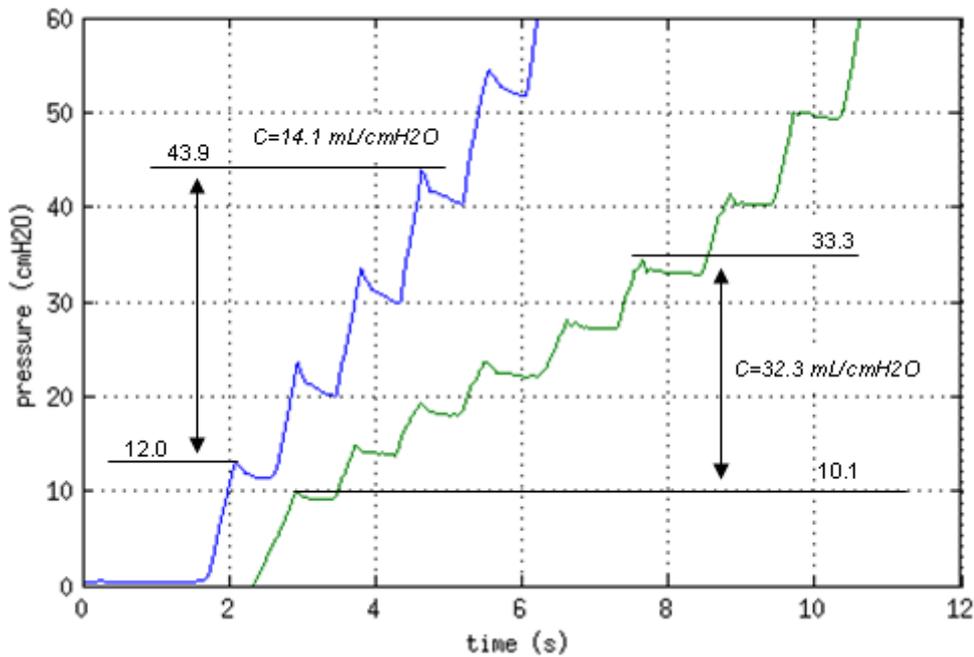

Figure 5 – Calibration of the lung models. Air was injected in the "lung" by a volumetric pump in steps of 150 mL.

In all cases, two oxygen fractions, measured at the exhaust of C2, were considered: 100% $O_2$ and less than 60% $O_2$. In many cases the lower $O_2$ fraction limit was below 50 %.

All curves presented are randomly chosen single breaths, not averages over many breaths.

Similar data is presented in Figure 7 for the 14 mL/cmH$_2$O "lung". In this case we considered limit PIP pressures of 30 or 40 cmH$_2$O, compatible with the 6 mL/kg maximum expectable volume intake and maximum safe PIP pressure.

All curves show a fast-initial pressure step of about 50% of the set PIP and then a slower convergence to the set PIP. This effect is stronger in the more compliant "lung", which accepts a larger tidal volume[†] for the same pressure. A comparison of the performance of several regulator models is presented in the Technical Annex. Models with larger flow capacity yield a slightly more "rectangular" pressure curve profile.

Owing to the simplicity of the timing electronics there is some influence of the I/E ratio on the breathing rate, slightly higher for I/E = 1/3, while the actual breathing rates are a bit faster than the nominal values, requiring further tuning of the circuit (see the Technical Annex).

---

[†] Inhaled volume in each inspiration.

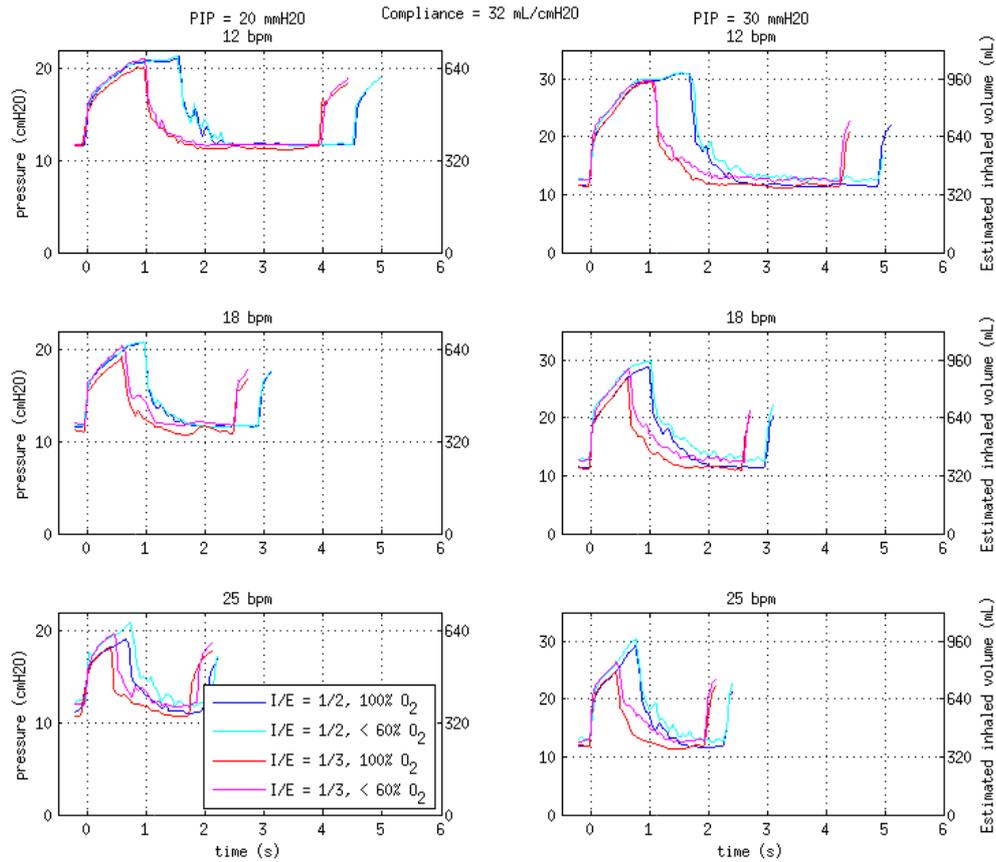

Figure 6 – Pressure profiles measured at the entrance of the "Y piece" as a function of time for breathing rates of approx. 12, 18 and 25 bpm, I/E ratios of 1/2 and 1/3, and lung compliance of 32 mL/cmH$_2$O. The PIP was set to 20 or 30 cmH$_2$O (left or right columns) and PEEP was set to 12 cmH$_2$O in all cases. Two fractions of exhaled O$_2$ were considered: 100% or less than 60% O$_2$/air fraction. The corresponding inhaled volume, estimated from the calibration, can be read on right hand side scale.

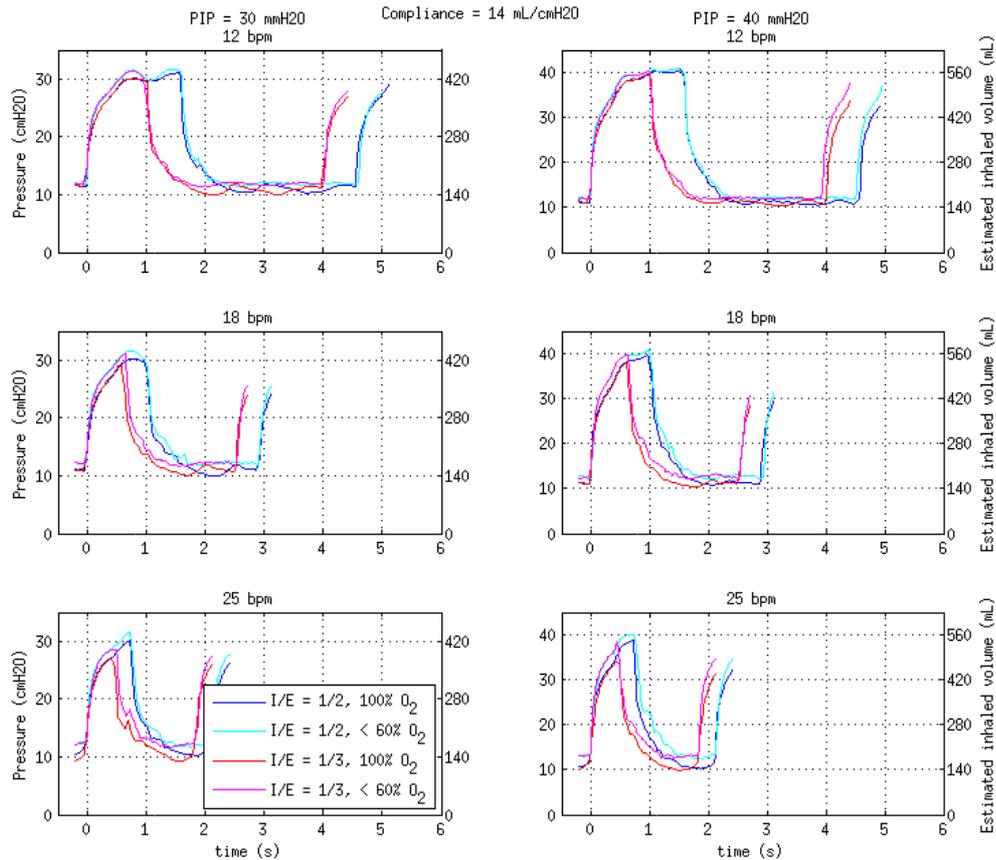

Figure 7 - Pressure profiles measured at the entrance of the "Y piece" as a function of time for breathing rates of approx. 12, 18 and 25 bpm, I/E ratios of 1/2 and 1/3, and lung compliance of 14 mL/cmH$_2$O. The PIP was set to 30 or 40 cmH$_2$O (left or right columns) and PEEP was set to 12 cmH$_2$O in all cases. Two fractions of exhaled O$_2$ were considered: 100% or less than 60% O$_2$/air fraction. The corresponding inhaled volume, estimated from the calibration, can be read on right hand side scale.

## *Durability*

At the moment of writing the ventilator had been subject to 35 full days of continuous operation in pure oxygen at a rate of 120 bpm, corresponding to 140 days of normal operation at 30 bpm, without any malfunction or unusual (ordinarily modest) warming.

## *Toxicity and biosafety*

### Chemical tests

The quality of the oxygen/air mixture that passes though the ventilator was determined by Gas Chromatography Mass Spectrometry (GC-MS) in order to look for potential volatile organic compounds that could arise from the materials used in its construction.

Prior to both the chemical and biological tests the inspiratory branch was ultrasonically washed in deionised water plus neutral detergent, thoroughly rinsed in deionised water and dried in an oven at 60 ºC for 12h.

A mixture of oxygen/air that passed through the equipment was fed through a sampling tube filled with sorbent (Supelco ORBO 43) for 1 hour. During this time the inspiration valve was cycled at a frequency of 120 cycles/minute.

After sampling, the tube was closed, kept in fridge and protected from light until the GC-MS analysis. The sample was extracted from the sorbent with toluene and analysed. A baseline sample of the oxygen/air fed to the ventilator was also likewise collected and analysed for comparison.

The toluene was obtained from Fisher Chemical (U.K) and used without further purification.

GC-MS analysis was performed in a GC-MS QP 2010 Plus from Shimadzu. Injections were performed using an AOC-5000 Auto Injector and a Supelco: SLB 5ms Fused silica capillary column, 60 m x 0.25mm ID, fused silica capillary, 0.25 µm. Data acquisition and analysis was performed with the software LabSolutions – GCMSolutions version 2.50 SU3.

Chromatographic conditions: injector temperature: 200ºC; detector temperature: 250ºC; interface temperature: 290ºC; oven temperature programme:130°C to 290°C at 4ºC for min (hold 20 minutes at the end); transporter gas: He; linear velocity: 35cm/sec; injection volume: 1µ; split ratio: 1.0.

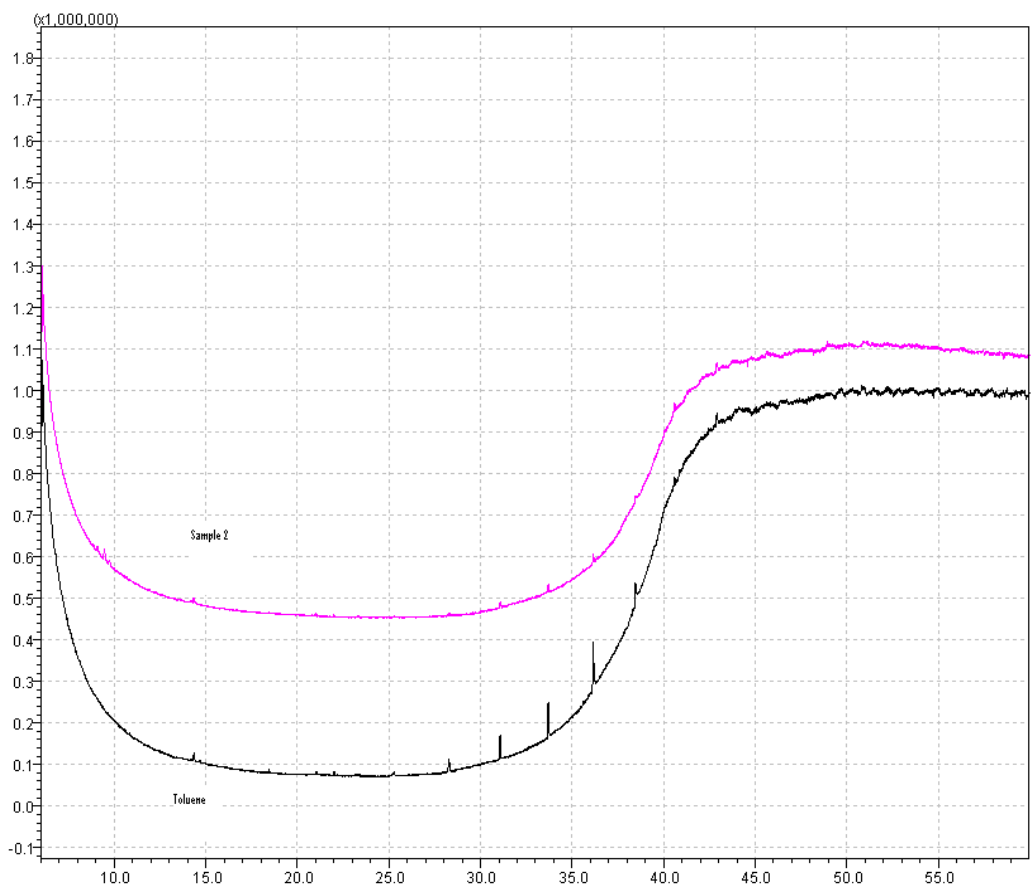

Figure 8 – GC-MS spectrum from the gas mixture fed to the device (black curve) and the gas mixture that passed through (pink curve). Horizontal scale: retention time (minutes). Vertical scale: Total ion current (arbitrary units).

The GC-MS analysis of both the gas mixture fed to the device (baseline) and of the gas mixture that passed through (sample) is presented in Figure 8, corresponding to a mass scan m/z between 30 and 300 amu.

It can be seen that the baseline actually shows more traces of extraneous compounds than the sample. This is attributed to residual compounds introduced from the gas distribution system, as the baseline was collected before the sample. Overall, no extraneous compounds were introduced by the device.

**Bacteriological tests**

Biological tests were also conducted on the oxygen/air mixture for the detection of bacteria and fungi.

Under the same conditions as for the chemical tests, a mixture of oxygen/air passing through the ventilator was bubbled for five minutes in water for injectables obtained from B-Braun Medical (Portugal) and used without further purification.

Samples were incubated in YEA (Yeast extract agar) and Rose-Bengal with chloramphenicol agar base (according ISO 6222) at 22ºC for 2 days and 36ºC for 3 days for detection of microorganisms and in MEA (malt extract agar) at 25ºC for 5 days for fungi (ISO 21527).

Both tests, for bacteria and fungi, were negative.

## Discussion

The results should be evaluated against the ventilation parameters relevant for the majority of COVID-19 patients:
- PIP not exceeding 30 $cmH_2O$ in most cases
- PEEP around 12 $cmH_2O$
- tidal volume not exceeding 6 mL/kg of ideal body weight
- breathing rates up to 25 bpm
- I/E ratio down to 1:3
- $FiO_2$ in the range 50% to 100%
- Lung compliance from 20 to 40 $mL/cmH_2O$

The set PIP is reached for most conditions tested, typically with an initial pressure step corresponding to at least 50% of the tidal volume. The deviation from the desirable rectangular pressure curve shape seems to be caused by the limited flow capability of the regulator. Higher I/E ratios or lower bpm improve the curve shape by providing more time for inspiration.

The non-conforming cases are confined to 25 bpm at I/E of 1/3. In the worst case a PIP of 25 $cmH_2O$ was reached when the set PIP was 30 $cmH_2O$.

The results are therefore conditioned by the flow characteristics of the pressure regulators. In the Technical Annex it can be found a comparison of several models of different capacities, but no substantial difference for this application was observed.

The admixture of air could lower the oxygen fraction in the exhaled gas to less than 60% in all cases, with only very slightly changes in the pressure curves. Actually this admixture helps to reach the set PIP for the most difficult settings without causing undue overpressure for the rest. However, owing to the simplicity of the device, the setting of a particular desired $O_2$ fraction takes some time, requiring successive small adjustments. Once set, the $O_2$ fraction was quite stable.

## Conclusion

The proposed emergency ventilator concept, implemented with a small number of common industrial components, aims to achieve volume and speed of production and independence from critical components, while in reasonable fulfillment of the clinical requirements for pressure-controlled continuous mandatory ventilation mode (PC-CMV), which may prove to be helpful on severe COVID-19 patients in conditions in which standard ventilators are unavailable.

The low cost of the device may extend its usefulness to lower or middle income countries.

The durability of the device was tested in pure oxygen with a number of cycles equivalent to at least 140 days of normal operation, without any adverse effect detected. These tests are continuing.

The toxicity of the materials present in the pressure regulator and the electrovalves was evaluated by gas chromatography, being found that no extraneous chemical compounds were introduced in the oxygen stream from the ventilator.

A biosafety test concluded that no bacteria or fungi could be cultivated from the oxygen stream.

From an industrial point of view we don't see any clear obstacle to the immediate production of this device in large scale at distributed sites around the World.

## Acknowledgement

To Estado Maior General das Forças Armadas (EMGFA) and Hospital das Forças Armadas (HFAR), namely Vice-Almirante Henrique Gouveia e Melo, Brigadeiro-General João Jácome de Castro, Coronel Médico Rui Teixeira de Sousa and Major Médico Ricardo Miguel Mimoso Ferreira of the Portuguese Armed Forces for their interest and support of this research.

We thank Olga Calado and the Microbiology Unit, Biocant, Center of Neurocience and cell Biology, for the biological tests and Rui Manadas from UCQfarma, Faculdade de Farmácia da Universidade de Coimbra for the GC-MS analysis.

To the companies
- Chamagás GALP
- Gavedra ([www.gavedra.pt](www.gavedra.pt))
- Intersurgical Portugal ([pt.intersurgical.com](pt.intersurgical.com))
- Refrimondego
- Tucab ([www.tucab.pt](www.tucab.pt))

for the kind donation of components.

To the companies José Garcia Lda and Ventiplast ([www.ventiplast.com](www.ventiplast.com)) to the special attention given to the production of parts for this project and to HACCPEMI Unip. Lda for the special availability of materials.

To the Pediatric Intensive Care Unit, Pediatrics Department, Hospital de Santa Maria, Centro Hospitalar Universitário Lisboa Norte, Lisboa, Portugal for providing test devices.

For the material support of UNIDEMI, the Universidade NOVA de Lisboa research centre in the field of Mechanical and Industrial Engineering.

To the wonderfully creative [https://www.helpfulengineering.org/](https://www.helpfulengineering.org/) community.

# Technical annex

## Mechanical

The main idea behind this project was to find in the local market/industry all the components, both in immediate availability and in considerable quantity. Other important issue was the cost, to allow the use of the device in lower income regions.

Hospital gas distribution lines are similar to the hydrocarbon supply lines at apartment buildings for example. Main distribution line pressure ranges between 4 and 5 bar, final pressure of some tens of mbar, flow rates higher than 20 L/min. Tubes and connectors are also suitable for both applications.

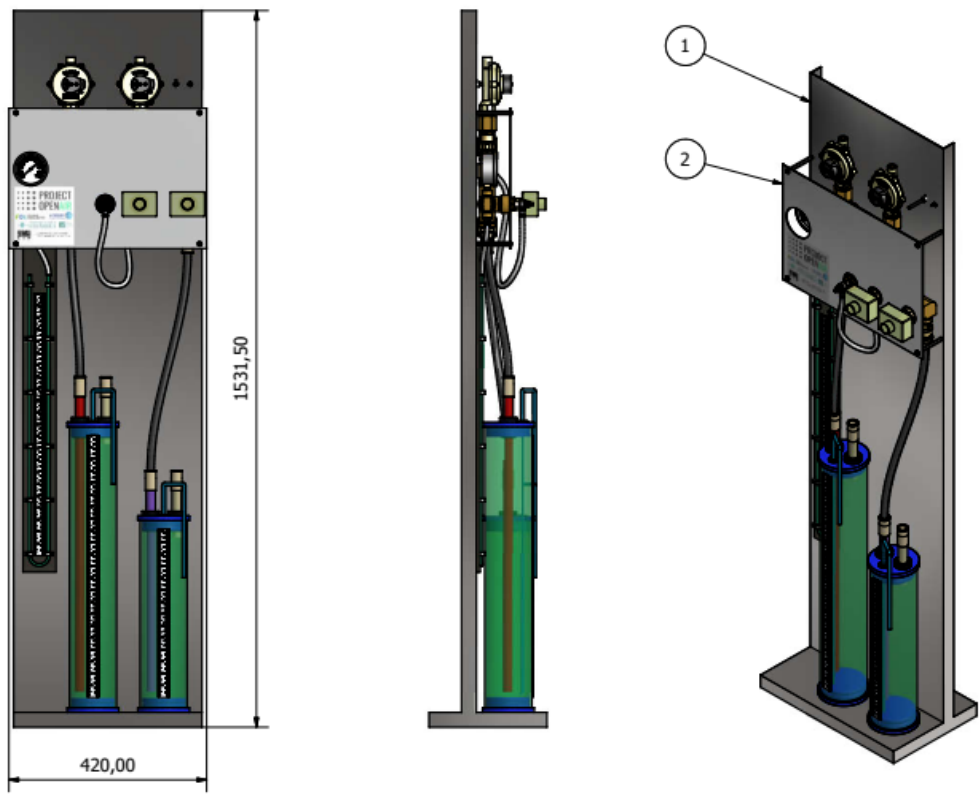

Figure A – Prototype views and external dimensions.

The main lines in the system were done with ¾" threaded connectors. When needed or convenient, reductions and fast connectors were also used. The commercial medical breathing systems standard connections are the 22F, so the system interface connectors should be 22M. Hospital main gas distribution systems use different kinds of standard connectors mostly correlated with country. In Portugal are used lockable AFNOR ones, which respect the standard NF S 90-116.

In figure A it is shown the complete system, where it is possible to identify all parts and be aware of the dimensions. The system is assembled in a stainless steel panel (1) and a metallic front plate (2) is used to block all the electric connections from user access. It is possible to add wheels for easier movement.

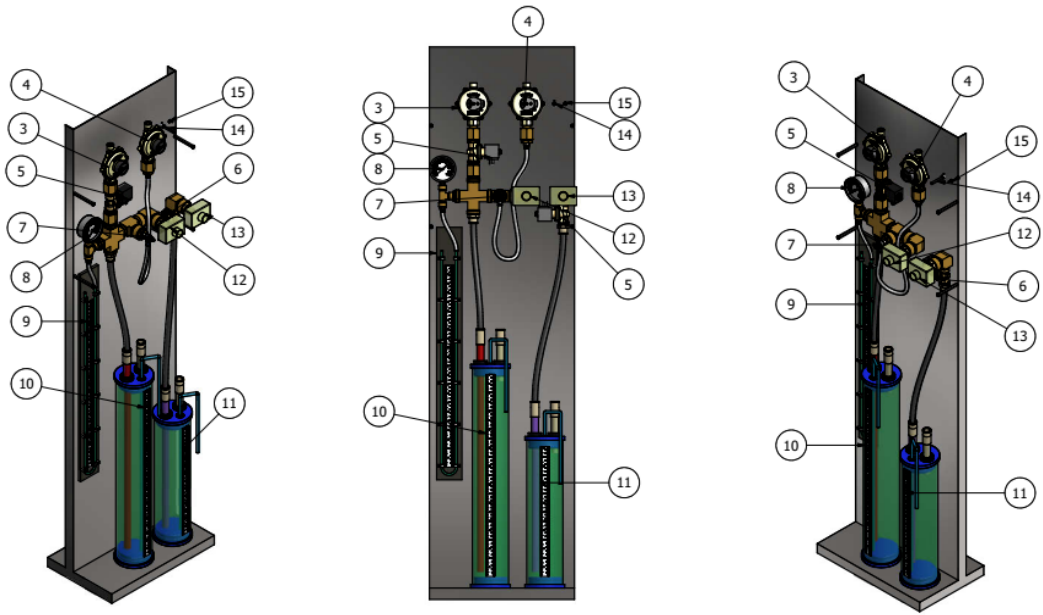

Figure B – Main parts identification.

As mentioned above, Oxygen and Medicinal Air are delivery around 4 bar. To reduce this pressure to PIP levels (lower than 45 mbar) were chosen low pressure regulators (3, 4). In the market it is possible to find a large number of suppliers from all around the World. It is preferable to consider a flow rate higher than 20 L/min or 5 kg/h. Some models required shortening the internal spring in order to reach down to 20 mbar of PIP pressure, for instance, by safely attaching together two coils of the spring.

Most of the rubbers used as membranes or seal O-rings are suitable for both gases. However a more careful check and test must be carried out to assure a safe choice.

The e-valves (5, 6) controlled by the timer (see electrical section) define the inspiration and expiration periods. Same as the low pressure regulators, there is an ample choice of solutions in the market. The valves should be able to draw the needed flow and allow opening pressures up to at least 100 mbar. If possible, the PEEP valve should be normally open so that in case of power failure the patient is depressurized to the PEEP pressure.

A needle valve (7) defines the Oxygen fraction in the mixture. The required Medicinal Air flow is low and as consequence it is possible to apply small section valve, tubes and connectors (e.g. 6 mm diameter).

To measure PIP and PEEP pressures we implemented two solutions. A mechanical manometer (8), optional since it may be not easy to find everywhere and/or in the needed quantities. A more universal solution is the water column manometer (9) that can be easily done with a flexible tube with enough transparency to check the water levels. Using any kind of ruler/meter it is very simple and accurate to determine the pressures.

To prevent any accidental excessive overpressures in the patient tubing, a safety column (10) is introduced. A large number of commercial safety valves suitable for that function can be found easily in the market, but unfortunately not in large quantities and within short production times. The PEEP column (11) defines the overpressure of the expiration branch. Commercial PEEP valves could be applied, but during pandemic times they could become out of stock. Both columns could be equipped with a ruler and the input tube could be adjusted by more than 20 cm in height. That way keeping the water level constant it is possible to adjust PIP safety and PEEP by 20 mbar. It is important to use corrugated tubes to help on this movement/adjust. To achieve pressures below 20 mbar in PIP safety column is necessary to reduce the water level. For more details on both columns see figures C and D.

Both inspiration and expiration branches "ends" with commercial bacteriologic filters (12, 13) and standard medical breathing tubes could be applied. The rotary switch (14) selects the breathing rates and can also select for permanent expiration or inspiration. The inspiration/expiration rates are select by the switch (15). For more detailed explanation on these switches see the electrical section.

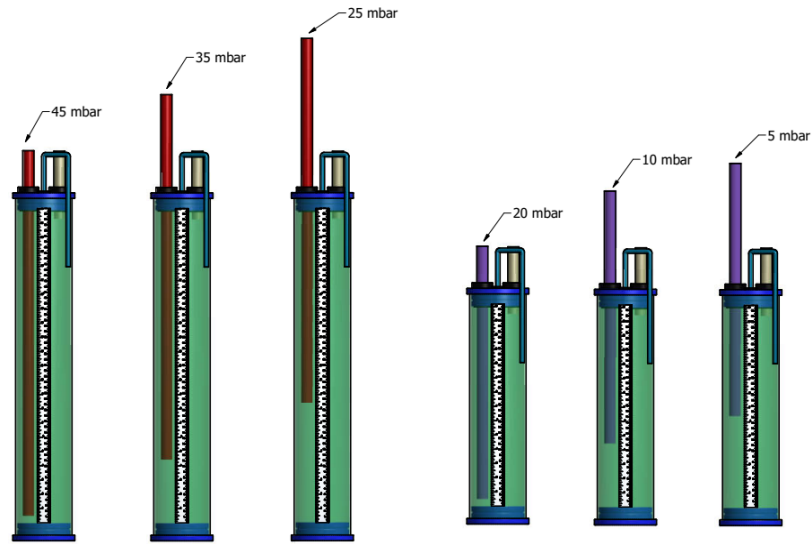

Figure C – Schematic explanation for pressure adjusts in PIP safety and PEEP columns.

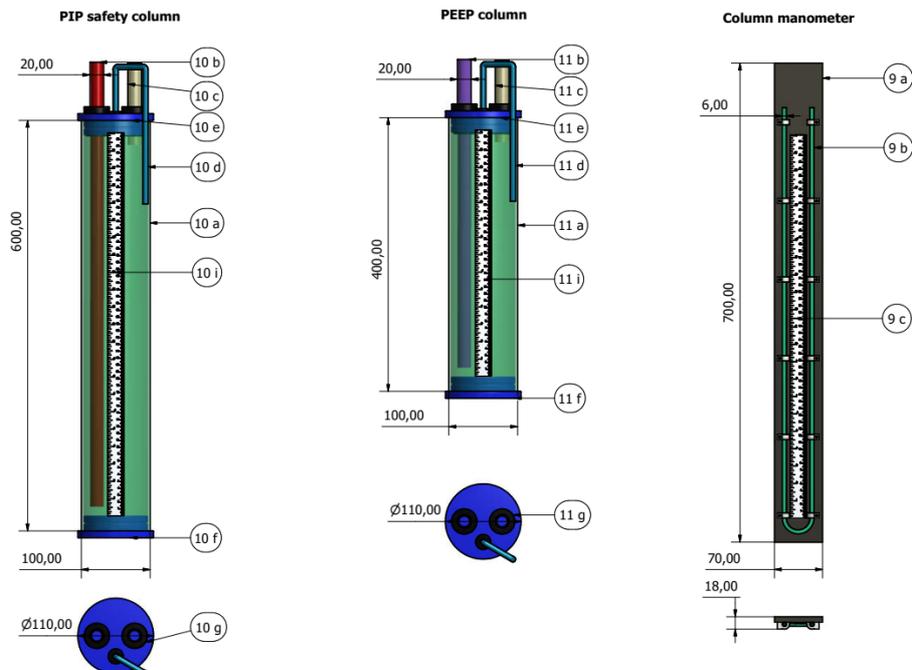

Figure D – Right - PIP safety column. Centre - PEEP Column. Left - U tube manometer.

PIP safety column parts identification (same is valid for PEEP column). The outer tube (10 a) has to be sufficiently transparent to check the water levels and input tube (10 b) position. The exhaust tube (10 c) should not be connected tightly to the hospital's ventilation main conduit as this may perturb the PEEP pressure. A small section flexible pipe (10 d) can be used to adjust the water levels. Any kind of ruler/scale (10 i) could be fixed to the outer tube for pressure estimation. The caps (10 e, 10 f) should be easily removed for column cleaning and sterilization. To assure enough tightness rubber O-rings (10 g) were applied.

The U–tube water manometer is just a flexible pipe (9 b) with enough transparency, a ruler/scale (9 c) and a base plate (9 a).

### Cleaning

Before assembly, all components should be cleaned ultrasonically with detergent and water, rinsed thoroughly with deionized water and sterilized with isopropyl alcohol (IPA) or ethyl alcohol at 70% v/v.

## Electrical

### Timer

The e-valves are commanded by a very simple timer based only on common and ubiquitous components, represented in Figure E.

The rotary switch S1 selects standard breathing rates of 12, 15, 18, 21 or 25 bpm (breathings per minute). Other values can be obtained by using more positions or it can be replaced by a potentiometer for continuous adjustment. Beware that in the later case some form of wiper position calibration or rate measurement will be needed.

The S1 switch also provides permanent expiration (off) and permanent inspiration positions.

If the rotary switch is difficult to source it can be replaced by independent switches in parallel. There is no major problem in selecting several bpm at the same time, except that the frequency will increase.

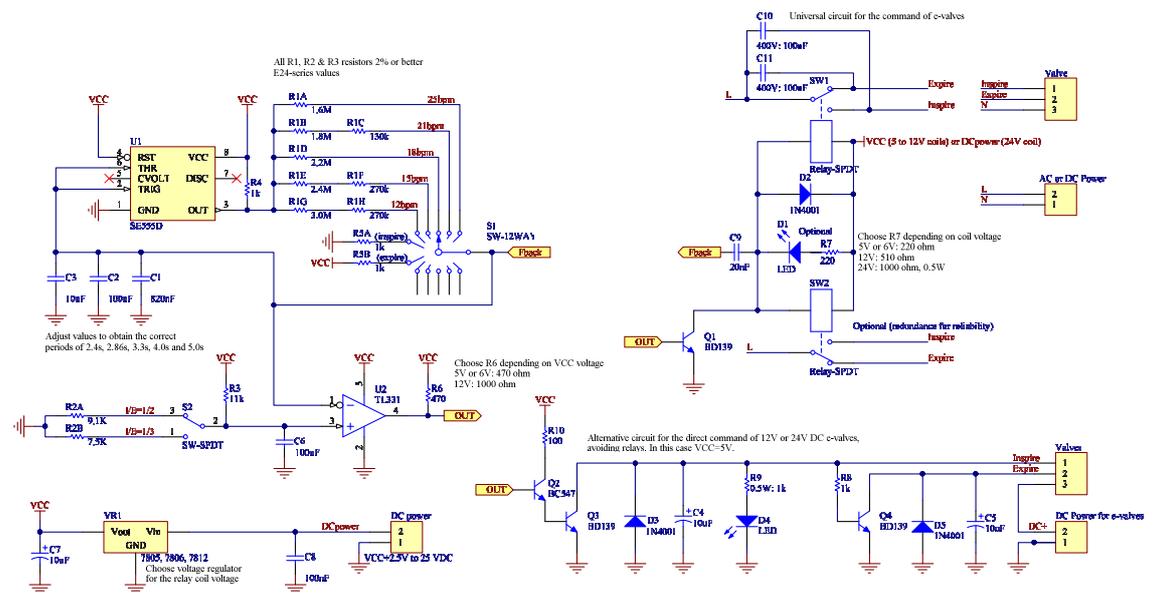

Figure E – Schematic representation of the timer circuit.

The inspiration/expiration time ratios are set in switch S2, for the most common fixed values of 1/2 (33% duty cycle) or 1/3 (25% duty cycle).

The electronically measured values of the respiratory rate and the I/E ratio are reported in Table 1 and compared with the nominal values, showing relative deviations of only a few percent. These deviations are caused by the finite precision of the values of the electronic components.

Table 1 - Measured values of respiratory rate (bpm) and ratio of inspiration time to expiration time (I/E).

| breathing rate (bpm) | | | I/E (%) | | | | | |
|---|---|---|---|---|---|---|---|---|
| nominal | measured | *% deviation* | nominal | measured | *% deviation* | nominal | measured | *% deviation* |
| 25 | 24.65 | *1.38%* | 75% | 78% | *-4.00%* | 66.6% | 63% | *5.41%* |
| 21 | 20.70 | *1.41%* | 75% | 77% | *-2.67%* | 66.6% | 63% | *5.41%* |
| 18 | 18.22 | *-1.24%* | 75% | 77% | *-2.67%* | 66.6% | 63% | *5.41%* |
| 15 | 15.04 | *-0.26%* | 75% | 77% | *-2.67%* | 66.6% | 63% | *5.41%* |
| 12 | 12.31 | *-2.59%* | 75% | 77% | *-2.67%* | 66.6% | 63% | *5.41%* |

The e-valves are driven by one or (optionally, for reliability) two SPDT 5V relays via the driver Q1.

For increased reliability the relays could be replaced by solid-state switches. For DC valves by power transistors driven in opposition (shown in Figure E), or, for AC valves, by triacs. Mind that in the later case there will be no galvanic insulation between the timer and the AC power line, so proper care should be exerted.

Alternatively, an industrial sequencer can be used, if available.

## *Apnea and power loss alarm*

The static pressure alarm detects the immobilization of the water manometer, indicative of a non-breathing situation (apnea). If the power is lost the alarm will sound for more than 1 minute.

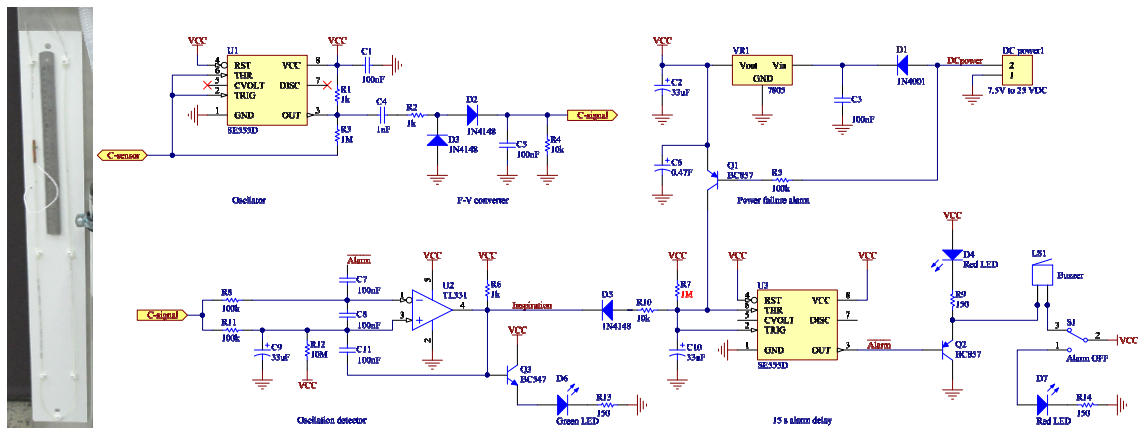

Figure F – Schematic representation of the apnea and power loss alarms and (left panel) position of the C-sensor as a metal electrode wrapped around the water manometer.

The subcircuit in the upper left corner should be mounted close to the metal electrode placed around the water manometer. It senses the presence/absence of water on that region of the tube, which varies the capacity seen by the oscillator and its frequency. If this oscillation is absent for longer than 15s the alarm will sound.

The alarm can be silenced, with indicator.

## Pneumatic

### *Comparison of different gas pressure regulators*

Several pressure regulators were compared in the setup described in [1], being the curves of pressure as a function of time presented in Figure F. The comparison was made at two PIP in the extreme ranges of normal operation: 45 and 20 cmH$_2$O. PEEP was kept fixed close to 12 cmH$_2$O. A long inspiration phase was chosen, corresponding to 12 bpm, I/E=1/2, because for shorter inspiration times the curves will be the same up to the time they are interrupted. The "lung" compliance was 36 mL/cmH$_2$O, which is less favorable (larger intake volume) for this purpose than stiffer lungs.

It can be seen that for PIP of 20 cmH$_2$O there is little difference in the curve shape between the five regulators tested. For PIP of 45 cmH$_2$O, corresponding to a very considerable intake volume of (45-12)×36 = 1.2 L, the regulators with more mass flow capacity, 7 or 10 kg/h yield a pressure curve a little closer to the ideal rectangular shape.

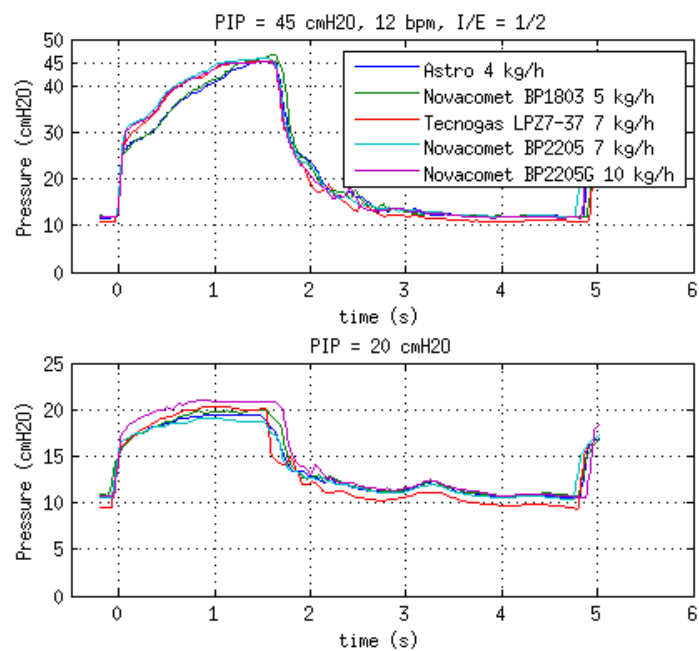

Figure F – Comparison of several types of common gas pressure regulators for a set PIP of 40 cmH2O or 20 cmH20. All (pressure vs. time) curves were taken at 12 bpm, I/E ratio = ½. The "lung" compliance was 36 mL/cmH$_2$O.